\documentstyle[12pt]{article}
\begin{document}
\newcommand{\bfig}{\begin{center}\begin{picture}}
\newcommand{\efig}[1]{\end{picture}\\{\small #1}\end{center}}
\newcommand{\flin}[2]{\ArrowLine(#1)(#2)}
\newcommand{\wlin}[2]{\DashLine(#1)(#2){2.5}}
\newcommand{\zlin}[2]{\DashLine(#1)(#2){5}}
\newcommand{\glin}[3]{\Photon(#1)(#2){2}{#3}}
\newcommand{\lin}[2]{\Line(#1)(#2)}
\newcommand{\sof}{\SetOffset}
\newcommand{\bmip}[2]{\begin{minipage}[t]{#1pt}\bfig(#1,#2)}
\newcommand{\emip}[1]{\efig{#1}\end{minipage}}
\newcommand{\putk}[2]{\Text(#1)[r]{$p_{#2}$}}
\newcommand{\putp}[2]{\Text(#1)[l]{$p_{#2}$}}
\newcommand{\bq}{\begin{equation}}
\newcommand{\eq}{\end{equation}}
\newcommand{\bqa}{\begin{eqnarray}}
\newcommand{\eqa}{\end{eqnarray}}
\newcommand{\nl}{\nonumber \\}
\newcommand{\eqn}[1]{eq. (\ref{#1})}
\newcommand{\eqs}[1]{eqs. (\ref{#1})}
\newcommand{\ibidem}{{\it ibidem\/},}
\newcommand{\vpb}{}

\title{
\vspace{-4cm}
\begin{flushright}
{\large  CERN-TH/98-221}\\ 
{\large  DEMO-HEP/98-02} 
\end{flushright}
\vspace{1.5cm}
{\tt WEXTER} and {\tt ERAFITTER}: two programs to fit $M_W$
at LEP2 using the best measurable kinematical variables}
\author{
        {\bf F. A. Berends}\\
        Instituut Lorentz, University of Leiden, P. O. Box 9506,\\
        2300 RA Leiden, The Netherlands\\\\
        {\bf C. G. Papadopoulos}\\
        Institute of Nuclear Physics, NCSR `Democritos',\\
        15310 Athens, Greece\\\\
          and\\\\
        {\bf R. Pittau}\\
        Theoretical Physics Division, CERN\\
        CH -1211 Geneva 23, Switzerland}
\date{}
\maketitle
\thispagestyle{empty}
\begin{abstract}
In this paper, we present two programs to fit $M_W$ at LEP2 
using the best measurable kinematical variables.
The theoretical probabilities of observing the final-state  
kinematical configurations are computed by integrating over the quantities
that are not well measured. Therefore, an
event-by-event kinematical reconstruction is avoided.  
$M_W$ is then determined through a maximum likelihood fit. 
\end{abstract}
\vspace{1cm}
\begin{flushleft}
{\large  CERN-TH/98-221}\\ 
July 1998
\end{flushleft}
\clearpage
\section{Introduction}
Two methods are mainly used at LEP2 to extract the $W$ mass:
the threshold method and the direct reconstruction technique \cite{lep2}.
In the first case the total $W^+ W^-$ cross section is measured near
threshold (161 GeV), where the sensitivity to $M_W$ is stronger, and
plotted as a function of the $W$ mass.
Conversely, the direct reconstruction method is applied at higher energies
and  requires two steps:\\

\noindent 1. From the experimental data the invariant mass distribution
  $\frac{d\sigma}{dM}$ is reconstructed. To improve the mass 
  resolution, a constrained fit is usually performed event by
  event, assuming no initial-state radiation (ISR) and equality between  
  the invariant masses coming from different $W$'s.\\

\noindent 2. The experimental distribution is compared with the
  theoretical prediction for $\frac{d\sigma}{dM}$   
  and, after Monte Carlo corrections, a reconstructed $W$ mass 
  $M_R$ is extracted, with an error $\Delta M_R$.\\

Recently, a new method has been proposed \cite{plb}
(direct fit method), in which only the best 
measured quantities are used to extract the $W$ mass. The idea is simple.
Given a set of well measured quantities $\{\Phi\}$ one 
computes, event by event, the theoretical probability ${\cal P}_i$ of
getting the observed set of values $\{\Phi_i\}$ for $\{\Phi\}$. 
Since this is a function of $M_W$,
given $N$ observed events, the logarithm of the likelihood function
($ L= \prod_i {\cal P}_i$)
is distributed, for large $N$, as a quadratic function of $M_W$.
A parabola can then be fitted, from which the reconstructed $W$ mass
$M_R$ is obtained with an error $\Delta M_R$.

Although one can always consider more sets $\{\Phi\}$, 
the following choices seem reasonable, in practice, for different
four-fermion final states \cite{plb,oxf}:
\begin{itemize}
\item[] Semileptonic case: $\ell_3 \nu_4 q_5 q_6 $
\begin{itemize}
  \item[1.] $\{\Phi\}= \{E_3, \Omega_3, \Omega_5, \Omega_6,
                      E_h\}$, where $E_h$ is the total energy of the two jets. 
  \item[2.] $\{\Phi\}= \{E_3, \Omega_3, \Omega_5, \Omega_6 \}$.  
\end{itemize}
\item[] Purely hadronic case: $q_3 q_4 q_5 q_6$
\begin{itemize}
  \item[3.] $\{\Phi\}= \{\Omega_3, \Omega_4, \Omega_5, \Omega_6 \}$.
\end{itemize}
\item[] Purely leptonic case: $\ell_3 \nu_4 \nu_5 \ell_6 $
\begin{itemize}
  \item[4.] $\{\Phi\}= \{E_3, \Omega_3,
                         E_6, \Omega_6 \}$.
\end{itemize}
\end{itemize}
In this paper, we explicitly give all the formulae needed to compute
the probabilities referring to the above four cases and present two
{\tt FORTRAN} programs ({\tt WEXTER} and {\tt ERAFITTER}) to extract 
$M_R$ from the LEP2 data using the direct fit method.
The two programs have been developed in parallel and continuously cross
checked. For this reason we chose to present them in a common paper.
Also, we decided to put emphasis on the description of the common
algorithms and on a general illustration of the codes, 
skipping the most technical details. Further information 
is available directly from the authors.

\section{Theory}
Let us start from the case when the analysed events
belong to only one particular class of four-fermion final states  
(for example $\mu^- \bar \nu_\mu q_1 \bar q_2$).

The main problem is computing, for each given event $i$, the probability 
${\cal P}(\{\Phi_i\},M_W)$ of measuring the observed values $\{\Phi_i\}$ 
for the set $\{\Phi\}$. This is a function of $M_W$ and reads
\bqa
 {\cal P}(\{\Phi_i\},M_W)= \frac{1}{\sigma_{tot}}\, 
\frac{d \sigma}{d \{\Phi_i\}}\,,
\eqa
where $\frac{d \sigma}{d \{\Phi_i\}}$ and $\sigma_{tot}$ 
are the differential and the total (namely integrated over the 
whole fiducial volume) cross sections for the process 
under study, respectively.
Once $\sigma_{tot}$ is known, the computation of 
$\frac{d \sigma}{d \{\Phi_i\}}\,$ for the four cases 
listed in the introduction is needed. Then, a likelihood function may be
constructed as follows: 
\bqa
L= \prod_{i} {\cal P}(\{\Phi_i\},M_W)\,.
\eqa
For large numbers of events, $L$ is distributed as a Gaussian. Therefore,
a quadratic fit of the form $Y= a\,X^2+b\,X+c$, 
with $Y= \log L$ and $X=M_W$, can be performed. 
The maximum of the fitted parabola ($-\frac{b}{2 a}$) gives  
$M_R$, while the statistical error given by the set of 
analysed events is $\sqrt{-\frac{1}{2 a}}$. 

The above procedure can be easily generalized to the case when 
the analysed events belong to different processes. In principle, 
this allows us to extract $M_R$, using together all LEP2 events
\footnote{Notice that, since the differential 
cross sections differ, also CP-conjugate processes such as 
$\mu^- \bar \nu_\mu q_1 \bar q_2$
and $\mu^+ \nu_\mu \bar q_1 q_2$ must be considered separately.}.
If the events refer to $m$ different processes, the likelihood
function is given by
\bqa
&& L                       = \prod_{i,j} {\cal P}^{(j)}(\{\Phi_i\},M_W) \nl
&& {\cal P}^{(j)}(\{\Phi_i\},M_W) = \frac{1}{\sigma^{(j)}_{tot}}\, 
                            \frac{d  \sigma^{(j)}}{d \{\Phi_i\}}\,, 
\eqa
where $j= 1:m$ labels the $m$ different processes.

Therefore, the most general formula for $\log L$ reads
\bqa
\log L= \sum_{j=1}^{m} \left( \sum_{i=1}^{N_j} 
\log \left( \frac{d  \sigma^{(j)}}{d \{\Phi_i\}} \right)
 -N_j \log \sigma^{(j)}_{tot} \right)\,,
\label{eqp}
\eqa
where $N_j$ is the number of analysed events for the $j^{th}$ process.

In the following, we list the algorithms needed to compute  
$\frac{d  \sigma^{(j)}}{d \{\Phi_i\}}$ for the four cases given
in the introduction. Due to the ISR, the sum of the two
incoming momenta reads
\bqa
 P = p_1+p_2= (E,p,0,0)\,,~~\,{\rm with}~~E= \frac{\sqrt{s}}{2} (x_1+x_2)\,,
                         ~p= \frac{\sqrt{s}}{2} (x_1-x_2)\,,
\label{Ep}
\eqa
where $x_{1,2}$ are the fractions of energy left to $e^{\pm}$ 
after QED radiation, and 
\bqa
 (p_1+p_2)^2 \equiv \hat{s}= x_1 x_2 s\,.
\eqa
\paragraph{Purely hadronic case\\}
This is the simplest case because 
$\{\Phi\}=\{\Omega_3, \Omega_4, \Omega_5, \Omega_6 \}$
is 8-dimensional. Then, when neglecting ISR, the kinematics 
of the event is completely determined.
The inclusion of ISR would instead imply two integrations over $x_1$ and $x_2$.

The algorithm is as follows.
First one generates $x_1$ and $x_2$ between $0$ and $1$, and computes 
$E$ and $p$ in eq. (\ref{Ep}).
Then, after parametrizing the final-state momenta as
\bqa
&& p_i= E_i\,(1,c_{\theta_i},s_{\theta_i} c_{\phi_i}
           ,s_{\theta_i} s_{\phi_i}) \nl
&& i= 3:6\,,~~~~c_{\theta_i} \equiv \cos{\theta_i}\,,
            ~~~~s_{\theta_i} \equiv \sin{\theta_i}\,,~~~{\rm etc.}\,, 
\eqa
the four unknown energies $E_i$ 
are found by using the energy-momentum conservation constraints: 
\bqa
&&\Delta
\cdot
\left( \begin{array}{l}
         E_3 \\
         E_4 \\
         E_5 \\
         E_6 
       \end{array}
\right) =
\left( \begin{array}{c}
         E \\
         p \\
         0 \\
         0 
       \end{array}
\right)\,\nl
&&\Delta= 
\left( \begin{array}{cccc}
         1           &1           &1           & 1            \\
         c_{\theta_3}&c_{\theta_4}&c_{\theta_5}&c_{\theta_6}  \\
         s_{\theta_3}c_{\phi_3}   &s_{\theta_4}c_{\phi_4}   & 
         s_{\theta_5}c_{\phi_5}   &s_{\theta_6}c_{\phi_6}     \\
         s_{\theta_3}s_{\phi_3}   &s_{\theta_4}s_{\phi_4}   &
         s_{\theta_5}s_{\phi_5}   &s_{\theta_6}s_{\phi_6} 
       \end{array}
\right) \,.
\eqa
For some values of $x_1$ and $x_2$, the above system may give
unphysical negative $E_i$. Such configurations must off course be discarded.

The matrix element squared $|M|^2$ can then be computed by
using the reconstructed momenta, so that the kernel cross section, 
to be convoluted with the ISR structure functions, is
\bqa
d \hat{\sigma}(x_1,x_2)= \frac{|M|^2}{16} \delta^4(P-p_3-p_4-p_5-p_6) 
          \prod_{i=3}^{6} E_i\,dE_i\,d\Omega_i \,,
\eqa
where $(2 \pi)^8$ and the flux factor have been included in the definition
of $|M|^2$.
Since
\bqa
\int dE_3\,dE_4\,dE_5\,dE_6\, \delta^4(P-p_3-p_4-p_5-p_6)= 
\frac{1}{{\rm det} \Delta}\,,
\eqa
the final answer reads
\bqa
\frac{d  \hat{\sigma}(x_1,x_2)}{d \{\Phi_i\}} \equiv
\frac{d  \hat{\sigma}(x_1,x_2)}{d \Omega_3 d \Omega_4 d \Omega_5 d \Omega_6}=
\frac{E_3 E_4 E_5 E_6}{16\,{\rm det}\Delta}\,|M|^2\,.
\eqa 
\paragraph{Purely leptonic case\\}
The measured quantities are now
$\{\Phi\}= \{E_3, \Omega_3, E_6, \Omega_6 \}$.
An additional integration over $\Omega_4$ is required,
besides that over $x_1$ and $x_2$.
Therefore one first generates
$x_1,\,x_2,\,\Omega_4$ and then, from the on-shell condition
$0= (P-p_3-p_4-p_6)^2$, one gets $E_4$:
\bqa
&& E_4= \frac{1}{2}\,\frac{\hat{s}-2\,E_3\,(E-p\,c_3)
     -2\,E_6\,(E-p\,c_6)+2\,E_3\,E_6(1-\,c_{36})}
     {E-p\,c_4-E_3\,(1-c_{34})-E_6\,(1-c_{46})} \nl
&& c_i= c_{\theta_i}\,,~~~~~~~c_{ij}= \cos \angle {(p_i,p_j)}\,.
\eqa
The four-vector $p_5$ is given by
$p_5= P-p_3-p_4-p_6$; then, for any value of 
$x_1,\,x_2\,{\rm and}\,\,\Omega_4$, the final-state momenta are known
(as in the previous case, unphysical solutions must be explicitly discarded). 
The kernel multidifferential cross section is therefore 
\bqa
&& \frac{d  \hat{\sigma}(x_1,x_2)}{d \{\Phi_i\}} \equiv
\frac{d  \hat{\sigma}(x_1,x_2)}{dE_3  d\Omega_3  dE_6  d\Omega_6} \nl
&& = \int d\Omega_4 
\frac{E_3 E_4 E_6}{16\,|E-p\,c_4-E_3\,(1-c_{34})-E_6\,(1-c_{46})|}\,|M|^2\,.
\eqa
\paragraph{Semileptonic case without hadronic energy\\}
Now $\{\Phi\}= \{E_3, \Omega_3, \Omega_5, \Omega_6 \}$ and one integration
is needed to compute the kernel differential cross section. We chose to
integrate over $E_5$.
As usual, one first generates $x_1$ and $x_2$. Then, a bound for $E_5$ 
is found from the condition $(P-p_3-p_5)^2 \ge 0$:
\bqa
 0 \le E_5 \le \frac{1}{2} 
 \frac{\hat{s}-2\,E_3\,(E-p\,c_3)}{E-p\,c_5-E_3\,(1-c_{35})}\,.
\eqa
By generating $E_5$ in the above interval, computing $E_6$ from
the on-shell condition $(P-p_3-p_5-p_6)^2 = 0$:
\bqa
&& \!\!\!\!\!E_6= \frac{1}{2}\,\frac{\hat{s}-2\,E_3\,(E-p\,c_3)
     -2\,E_5\,(E-p\,c_5)+2\,E_3\,E_5(1-\,c_{35})}
     {E-p\,c_6-E_3\,(1-c_{36})-E_5\,(1-c_{56})}
\eqa
and discarding the unphysical solutions, the final-state momenta 
are reconstructed.
Finally, the kernel cross section reads
\bqa
&& \frac{d  \hat{\sigma}(x_1,x_2)}{d \{\Phi_i\}} \equiv
\frac{d  \hat{\sigma}(x_1,x_2)}{dE_3 d\Omega_3 d\Omega_5 d\Omega_6} \nl
&& = \int dE_5 
\frac{E_3 E_5 E_6}{16\,|E-p\,c_6-E_3\,(1-c_{36})-E_5\,(1-c_{56})|}\,|M|^2\,.
\eqa
\paragraph{Semileptonic case with hadronic energy\\}
For this case, $\{\Phi\}= \{E_3, \Omega_3, \Omega_5, \Omega_6,
                      E_h\}$, where $E_h= E_5 + E_6$.
The set $\{\Phi\}$ is 8-dimensional; therefore, by giving $x_1$
and $x_2$, the kinematics of the event is completely fixed.
The algorithm is as follows.
First one generates $x_1$ and $x_2$, then one computes $E_5$ and $E_6$
from the system
\bqa
\left\{ \begin{array}{l}
        E_h= E_5+E_6 \\
        (P-p_3-p_5-p_6)^2=0\,. 
        \end{array}
\right.
\eqa 
By imposing $E_{5,6}= \frac{E_h}{2} \pm \delta$, one finds the
following two solutions:
\bqa
&& \delta^{\pm}= \frac{-\beta \pm \sqrt{D}}{2\,\alpha}\,, \nl
&& D= \beta^2-4\,\alpha\,\gamma~~~~~~~~\alpha= -2\,(1- c_{56})\,, \nl
&& \beta= 2\,\left[ p\,(c_5-c_6)+E_3\,(c_{36}-c_{35}) \right]\,, \nl
&& \gamma= \hat{s}-2E_3\,(E-p\,c_3)- E_h(2E-p\,(c_5+c_6)) \nl
&&~~+E_3 E_h\,(2-c_{35}-c_{36})+\frac{E_h^2}{2}\,(1-c_{56})\,.
\eqa
Then $p_4= P-p_3-p_5-p_6$. Not always are both solutions physical.
The conditions to be fulfilled are
\bqa
\left\{ \begin{array}{l}
        | \delta | \le \frac{E_h}{2} \\
        D \ge 0 \\
        E_4 \ge 0\,. 
        \end{array}
\right.
\eqa
Finally, the kernel cross section is
\bqa
&& \frac{d \hat{\sigma}(x_1,x_2)}{d \{\Phi_i\}} \equiv
\frac{d \hat{\sigma}(x_1,x_2)}{dE_3 dE_h  d\Omega_3  d\Omega_5  d\Omega_6} \nl
&& = \int d\delta\, 
\frac{|M(\delta)|^2}{8}\,E_3\,\left( \frac{E_h^2}{4}-\delta^2 \right)
\, \delta\left((P-p_3-p_5-p_6)^2 \right) \nl
&& = F(\delta^+)+F(\delta^-)\,, \nl
&& F(\delta)= \frac{|M(\delta)|^2}{8}\,E_3\,
\left( \frac{E_h^2}{4}-\delta^2 \right)\,\frac{1}{|2\alpha\delta+\beta|}\,.
\eqa
\section{The program {\tt WEXTER}}
In this section, we present the first of the two programs that use
the described probabilities to extract $M_R$ from the LEP2 data.
The program {\tt WEXTER} consists of three parts: the evaluation 
of the matrix element, the computation of the relevant
differential cross sections, by integration over the unobserved
variables, and the fit to the likelihood curve, to extract $M_R$.

The knowledge of the total cross sections 
$\sigma^{(j)}_{tot}$ in eq. (\ref{eqp}) 
is required as an input. The needed $\sigma^{(j)}_{tot}$ 
can be computed once for all, for each value of $M_W$ used in the fit,
using, for example, {\tt EXCALIBUR} \cite{exca}.

After an initialization in {\tt SUBROUTINE SETPRO}, the matrix 
element evaluation is performed in {\tt SUBROUTINE MATRIX}
and {\tt SUBROUTINE DIAGA}.
The differential cross sections are evaluated in {\tt SUBROUTINE DIFF},
while all needed integrations and the fit are performed in the 
{\tt MAIN} of the program.

\subsection{The {\tt MAIN}}
In the {\tt MAIN}, the input file is read 
(see later for a detailed discussion). The input file must include
the name of the data file ({\tt DATANAME}) containing the events to be
fitted. In {\tt DATANAME}, all events must be given in terms of a complete set
of four-momenta readable with the following format:
\begin{verbatim}
       open (unit=2,file=DATANAME,status='old')
       read (2,60) lp,(p(0,kl),p(1,kl),p(2,kl),p(3,kl),kl= 3,6)
   60  format(i4/,(4d19.10))
\end{verbatim}
where {\tt lp} is a flag that defines the process and 
the array {\tt p(0:3,1:6)} contains the four-momenta.

The first index in {\tt p(0:3,1:6)} refers to the component 
while the second one labels
the particles: $1$ is the incoming $e^+$, $2$ the incoming $e^-$, while
$3$, $4$, $5$ and $6$ are the four outgoing fermions.

By convention, the beam is along the $x$-axis (component 1), with the incoming
$e^{+}$ along the positive values.

\par From the $i^{th}$ event given in the above form, the program reconstructs the 
values $\{\Phi_i\}$ for the set $\{\Phi\}$ of best measured variables
(see section 2 and ref. \cite{plb}), namely energies and
angles for charged leptons and solid angles for the quarks.
The set $\{\Phi\}$ is automatically determined according to the value 
of {\tt LFLAG} returned by {\tt SUBROUTINE SETPRO} (see next section).

Of course, the above input format is not suitable when analysing
real data: in that case the values of the measured variables
$\{\Phi_i\}$ should be directly
given as an input. This requires a trivial change in the reading
format of the program.
However, we chose to feed the program directly with the four-momenta
in order to facilitate Monte Carlo studies. 

Then, the probability ${\cal P}(\{\Phi_i\},M_W)$ of measuring the observed 
values $\{\Phi_i\}$ for the set $\{\Phi\}$ is computed, 
by a Monte Carlo integration over the initial-state
QED radiation  - implemented as in ref. \cite{iqed} - 
and, when necessary, over the unobserved quantities.

Finally, in the last part of the {\tt MAIN}, the quadratic fit
$Y= a\,X^2+b\,X+c$ described in the previous section is performed
to extract $M_R$ from the analysed events. Also the correlation matrix 
is computed, to estimate the error on the fit due to the Monte Carlo 
integration.

\subsection{The subroutines {\tt SETPRO}, {\tt MATRIX} and {\tt DIAGA}}
These three subroutines and the whole strategy for the computation 
of the matrix element are taken from {\tt EXCALIBUR} \cite{exca}, 
to which we refer for further details.

In the first part of {\tt SUBROUTINE SETPRO} the strong and the electroweak 
parameters used in the program are set.
They are $M_Z$ ({\tt ZMI}), $\Gamma_Z$ ({\tt WZI}),
$\sin^2 \theta_W$ ({\tt STH2}), $\alpha_{e.m.}$ ({\tt ALPHA}) and
$\Gamma_W$ ({\tt WWI}). Also $\alpha_s$ is an input. In the program,
two different $\alpha_s$'s are used. The first one ({\tt ALS}) controls the
coupling of the additional gluonic diagrams appearing 
in four-quark final states \cite{qcd} (setting {\tt ALS= 0} switches
off such diagrams). The second one ({\tt ALSN}) is used in the computation
of the so-called ``naive'' QCD factor \cite{qcdn}.

By default, the widths of the gauge bosons (set in {\tt FUNCTION CM2})
are taken to be fixed and different from zero also in the $t$-channel. 
This ensures QED gauge invariance of the results \cite{pask}, 
but induces a small shift in $M_W$. To compensate for it, the variable
{\tt LSHI} is introduced. When {\tt LSHI} is chosen to be 1 in the
input file, masses and widths are redefined as described in ref. \cite{shift}.
If {\tt LSHI = 0} such a redefinition is not performed.

Then, the four-fermion processes, chosen in the input file, are read and
the corresponding Feynman diagrams built up and printed in the output
file. In {\tt SUBROUTINE SETPRO}, the chosen processes are also classified 
according to four categories.
The variable that controls the classification is {\tt LFLAG}:
{\tt LFLAG= I (I= 1:4)} corresponds to the {\tt I}-th choice for
$\{\Phi\}$ described in the introduction. 

Finally, the right permutation of the four particles in the final state
is assigned to the variables {\tt N3, N4, N5} and {\tt N6}, for later 
use in {\tt SUBROUTINE DIFF}. This is relevant for leptonic and
semileptonic final states only. In fact, all four particles 
in fully hadronic final states are equivalent.

In {\tt SUBROUTINE MATRIX} and {\tt DIAGA} the matrix element 
squared is computed, using the helicity techniques 
described in ref. \cite{exca}.

\subsection{The subroutine {\tt DIFF}}
In {\tt SUBROUTINE DIFF} the differential cross section is computed
according to the value of {\tt LFLAG} given by the event.
The algorithms for the computation have already been  described in
section 2. {\tt SQJAC} is the value returned by the subroutine. It
is the product of the matrix element times the relevant Jacobian.
{\tt SUBROUTINE DIFF} is called from the {\tt MAIN} of the program,
where the numerical integration over ISR and not measured variables 
is performed.
\subsection{The input}
The meaning of the input parameters to be specified in 
order to run {\tt WEXTER} is the following:
\begin{itemize}
\item {\tt OUTPUTNAME (CHARACTER*15)}\\
The name of the output file.
\item {\tt DATANAME (CHARACTER*15)}\\
The name of the file containing the events.
\item {\tt NLP (INTEGER)}\\
The number of different processes contained in the file {\tt DATANAME}.
\item {\tt PAR(3,I) (CHARACTER*2)}\\
Produced fermion with label 3 for process {\tt I} 
( to be chosen among 
 '{\tt EL}', '{\tt NE}', '{\tt MU}', '{\tt NM}', '{\tt TA}',
 '{\tt NT}', '{\tt DQ}', '{\tt UQ}', '{\tt SQ}', '{\tt CQ}',
 '{\tt BQ}', '{\tt TQ}).
\item {\tt PAR(4,I) (CHARACTER*2)}\\
Produced antifermion with label 4 for process {\tt I}.
\item {\tt PAR(5,I) (CHARACTER*2)}\\
Produced fermion with label 5 for process {\tt I}.
\item {\tt PAR(6,I) (CHARACTER*2)}\\
Produced antifermion with label 6 for process {\tt I}.
\end{itemize}
The block of the previous four entries, specifying the
processes, should be repeated {\tt NLP} times ({\tt I= 1:NLP}).  
\begin{itemize}
\item {\tt NEV (INTEGER)}\\
The number of events in the file {\tt DATANAME} to be analysed for the fit.
\item {\tt NPF (INTEGER)}\\
The number of different values of $M_W$ used to 
fit the likelihood curve.
\item {\tt N (INTEGER)}\\
The total number of integration points.
\item {\tt KREL (INTEGER)}\\
It selects the Feynman diagrams. If {\tt KREL = 0} all possible Feynman diagrams
contributing to the chosen processes are taken into account. 
If {\tt KREL= 1}, only the {\tt CC03} diagrams leading to the reaction
$ e^+ e^- \to W^+ W^- $.
\item {\tt LQED (INTEGER)}\\
It includes ({\tt LQED= 1}) or excludes ({\tt LQED = 0}) ISR.
\item {\tt LCOUL (INTEGER)}\\
It includes ({\tt LCOUL= 1}) or excludes ({\tt LCOUL = 0}) the
Coulomb factor described in ref. \cite{coul}.
\item {\tt LQCD (INTEGER)}\\
It includes ({\tt LQCD= 1}) or excludes ({\tt LQCD = 0}) the
``naive'' QCD factor \cite{qcdn}.
\item {\tt LENER (INTEGER)}\\
For semileptonic processes, the sum of the hadronic energies 
is used ({\tt LENER= 1}) or not used ({\tt LENER = 0}) 
as an input. In other words, if {\tt LENER= 1},
a semileptonic process is classified with {\tt LFLAG= 1}.
Otherwise with {\tt LFLAG= 2}.
\item {\tt LFOLD (INTEGER)}\\
For semileptonic (hadronic) processes a 2-folding (24-folding) 
over all possible jet assignments is performed 
({\tt LFOLD= 1}) or not performed ({\tt LFOLD= 0}).
\item {\tt LSHI (INTEGER)}\\
A shift of masses and widths as described in ref. \cite{shift}
is performed ({\tt LSHI= 1}) or not performed ({\tt LSHI= 0}).
\item {\tt ROOTS (REAL*8)}\\
The total energy (in GeV) of the colliding $e^+$ and $e^-$.
\item {\tt X(I) (REAL*8)}\\
{\tt NPF} values of $M_W$ (in GeV) used to fit the likelihood curve
({\tt I= 1:NPF}).
\item {\tt SIG(I,J) (REAL*8)}\\
{\tt NPF} values ({\tt I= 1:NPF}) of the cross section integrated 
over the fiducial volume for the process labelled with {\tt J}.
The {\tt I}-th cross section must be computed with 
the {\tt I}-th value for $M_W$ ({\tt X(I)}). 
These and the following quantities are an input 
for {\tt WEXTER} and should be computed
once for all using, for example, {\tt EXCALIBUR} \cite{exca}.
\item {\tt DSIG(I,J) (REAL*8)}\\
The {\tt NPF} values ({\tt I= 1:NPF}) of the error 
corresponding to the above quantities.
\end{itemize}
The block of the previous two entries, specifying the
cross sections and their errors, should be repeated 
{\tt NLP} times ({\tt J= 1:NLP}).
\subsection{The output}
Since presenting a complete test run output
would require the specification of too long a list of numbers, 
such as the events contained
in the file {\tt DATANAME}, we decided not to include it here.
We just describe what a typical output file looks like.
After printing out information about the process and the parameters 
given in the input file, the program writes down, 
for each value of $M_W$ chosen for the fit, the following quantities:
\bqa
\sum_{j=1}^{m}\sum_{i=1}^{N_j} 
\log \left( \frac{d  \sigma^{(j)}}{d \{\Phi_i\}} \right)\,,~~~
\sigma_{tot}^{(j)}\,,~~~X= M_W\,,~~~Y= \log L\,.
\eqa
Then the final result of the fit is reported in the following form:
\begin{verbatim}
          FITTED MW WITH A 3 PARAM FIT (Y= a*X^2+b*X+c) :          

          wm=  0.803187D+02 +/- 0.503275D-01 +/- 0.750389D-02

          Chi^2/d.o.f. =  0.129242D+01

          a = -0.197406D+03
          b =  0.317107D+05
          c = -0.128700D+07
\end{verbatim}
where {\tt wm} is the reconstructed $W$ mass ($M_R$) and the first and
second errors are the statistical and the Monte Carlo errors,
respectively. The previous result has been obtained by analysing a
set of 1600 unweighted {\tt CC03} semileptonic events, produced by
{\tt EXCALIBUR}, with an input mass $M_W= 80.35$ GeV at $\sqrt{s}= 190$ GeV. 
This set of unweighted events, together with the input
file used to get the above output, are available, upon request,
from the authors. 
\section{The program {\tt ERAFITTER}}
The program provides all necessary elements to fit the $W$ mass and
it is based on {\tt ERATO} \cite{erato}. Below we give a brief
description of the program.

\subsection{The Computational Tree}

The program {\tt main.f} evaluates the differential cross section for 
the different cases described above. Phase-space generation proceeds
through the appropriate algorithms {\tt ALGO01} to {\tt ALGO04}. Then the
momentum assignment is used to  evaluate the matrix element through 
the routine {\tt MASTER}, which is extracted from {\tt ERATO}, and it is
specific for each selected channel, i.e. leptonic, semileptonic and hadronic.  
The total cross sections, needed for the evaluation of the likelihood function,
are calculated in the programs {\tt evud\_mass.f}, {\tt llll\_c2\_mass.f} and 
{\tt qqqq\_mass.f}. Finally {\tt fit2.f} collects all information and 
does the actual fit to extract the $W$ mass by using standard 
{\tt MINUIT} \cite{minuit} calling sequences.

In {\tt main.f}, the input data are read and all physical 
constants needed for the computation are defined. The output is the
value of $\sum_{i=1}^{N} \log(d\sigma/d\{\Phi_i\})$ for the various values
of $M_W$ specified by the user.

In the subroutines 
{\tt setc1}-{\tt setc4} the transformation from the 
Cartesian representation of the four momenta to the polar one, needed
to define the well-measured variables, is performed.

In the subroutines {\tt ALGO01}-{\tt ALGO04}, the generation of the 
full phase space, including, if specified, the ISR,
is performed, according to the discussion in section 2. Each subroutine
corresponds to the four cases defined in the introduction. 
Then the matrix element
is computed by the standard call of {\tt ERATO}. All needed routines
are incorporated in the file {\tt comrou.f}.

As far as the common blocks used in this code are concerned,
they are identical to the ones used by {\tt ERATO}, 
with the addition of {\tt VARI01}-{\tt VARI04}, which give
the values of the well measured variables in each case,
and {\tt masses}, where {\tt ndim} is the total number of $M_W$ values
used in the computation and {\tt dwmas} is an array containing these
values. Finally, {\tt iaprox} defines the approximation scheme 
according to which the $M_W$ dependence is taken into account.
If {\tt iaprox= 1} the $M_W$ dependence is computed 
by only considering the Breit-Wigner functions, 
\[ 1/((s_{12}-M_W^2)^2+M_W^2 \Gamma_W^2)\times 
   1/((s_{34}-M_W^2)^2+M_W^2 \Gamma_W^2) 
\,,\]
which is exact in the {\tt CC03} class of Feynman diagrams, and constitutes
a good approximation for {\tt CC10}, {\tt CC11} and, depending on the cuts,
for {\tt CC20} as well. On the other hand, if {\tt iaprox= 0} the
exact computation is performed. 
\subsection{Input description}
The input needed to run the {\tt main.f} code looks as follows: 
{\tt
\begin{verbatim}
3           !case 1=1b 2=1a 3=2 4=3 
190         !energy 
1 1600      !first and last generated event read 
0.3 10000   !derr,maxmc
9 79.73 79.93 80.03 80.13 80.23 80.33 80.43 80.53 80.73 
                                             !nim,dwmas
1           !isr
0 0 1       !ipro icoulomb iaprox
'/users/papadopo/Fortr/tmp/newfit/gen/run/evud/unw190100.data'
'lili100f'
'mass100f'
128.07 0.2310309 91.1888 2.4974 80.23 2.033   !input parameters
24           !# assignments (foldings) default =1 -> no folding
\end{verbatim}
}
\noindent The {\tt FORTRAN} variables corresponding to the above 
input file are:
\begin{itemize}
\item {\tt ial}\\ The flag defining the choice of the set $\{\Phi\}$,
 according to the order given in the introduction.
\item {\tt e0} \\ The collision energy.
\item {\tt nev1, nev2}\\ The first and last event read for analysis.
\item {\tt derr, maxmc}\\ The relative error required in the computation
of $d\sigma/d\Omega$ and the maximum number of MC iterations.
\item {\tt nim,dwmas(20)}\\ The number of and the actual values of $M_W$ used
in the calculation.
\item {\tt isr, ipro, icoulomb, iaprox} \\The flags for the ISR, the
actual process (i.e. $e^+e^-\to q_1\, \bar q_2\, q_3\, \bar q_4 $),
Coulomb correction and the approximation scheme as defined above.
\item {\tt pupu}\\ The input file where the events are stored.
\item {\tt pupu}\\ The output file containing the log-likelihood values
for each $M_W$.
\item {\tt pupu}\\ The output file containing the invariant masses of the
$W$ decay products.
\item {\tt ALPHA1, SINW2, ZMAS, ZGAMA, WMAS, WGAMA}\\ The physical constants
needed for the computation.
\item {\tt nfol}\\ The parameter controlling the number of foldings to be done,
{\tt nfol=2} for semileptonic and {\tt nfol=24} for hadronic. By folding we 
mean all possible different jet assignments.
\end{itemize}
\noindent For {\tt fit2.f} the first input file is:

\begin{verbatim}
inf_min_1: 
'../total/qqqq100'     !input file (tot_cs or lili)
'tot2'                 !output file (tot_cs_coef)
80.23                  !mass_0
0                      !dnorm
0                      !events 
1,5                    !ndat1-ndat2
0,1                    !option 0(2) ,ifirst 1(2)
\end{verbatim}

\noindent with the following correspondence to the {\tt FORTRAN} variables:

\begin{itemize}
\item{\tt FILNAM}\\
The input file including the total cross section for different values 
of the $W$ mass.
\item{\tt FILNAM}\\ 
The output file containing the coefficients of the fit of the total
cross section:
\[
\sigma=\mbox{par(1)}+\mbox{par(2)}\, M_W+\mbox{par(3)}\, M_W^2\,.
\]
\item{\tt rmax}\\
The assumed central value for the measured $M_W$, usually the input value
in a Monte Carlo simulation.
\item{\tt dnorm}\\
A normalization parameter (default = 0).
\item{\tt nev}\\
Not used.
\item{\tt ndat1, ndat2}\\
The first and the last rows to be read by the code from the
input file.
\item{\tt ioption, ifirst}\\
{\tt ioption=0} and {\tt ifirst=1}  in order to perform the
total cross section fit.
\end{itemize}

\noindent The second input file reads 

\begin{verbatim}
inf_min_2:
'lili100f'             !input file (tot_cs or lili)
'tot2'                 !output file (tot_cs_coef)
80.23                  !mass_0
0                      !dnorm
1600                   !events 
145 153                !ndat1-ndat2
2 2                    !option 0(2) ,ifirst 1(2)
\end{verbatim}

\noindent with

\begin{itemize}
\item{\tt FILNAM}\\
The input file including the log-likelihood values for each event and  
for the different values of the $W$ mass required.
\item{\tt FILNAM}\\ 
The output file containing the coefficients of the fit and the fitted $M_W$.
\item{\tt rmax}\\
The assumed central value for the measured $M_W$.
\item{\tt dnorm}\\
A normalization parameter (default = 0).
\item{\tt nev}\\
The number of events read by the {\tt main.f} code.
\item{\tt ndat1, ndat2}\\
The first and the last rows to be read by the code from the
input file.
\item{\tt ioption, ifirst}\\
{\tt ioption=2} and {\tt ifirst=2}  in order to perform the
fit to the logarithm of the likelihood function.
\end{itemize}

\noindent The first run performs the fit to the total cross section, 
whereas the second one uses the previous results to extract $M_R$, 
including the information from the differential
cross-section.

\subsection{The output}
Using the input files described above, we performed the fit to 
a selection of 1600 {\tt CC03} unweighted events produced by {\tt ERATO}
with ISR and $M_W=80.23$~GeV. Then a 24-folding was performed 
over all possible jet assignments. Moreover, the total cross section
has been evaluated using the same settings ({\tt CC03}, ISR).
The result of the {\tt MINUIT} program is as follows: 
\begin{verbatim}
**********
**    3 **MIGRAD
**********

MIGRAD MINIMIZATION HAS CONVERGED.

FCN=.3168413E-01 FROM MIGRAD STATUS=CONVERGED 
                                            13 CALLS  137 TOTAL
    EDM=   .57E-21  STRATEGY=1    ERROR MATRIX UNCERTAINTY= .0%

 EXT PARAMETER                             STEP         FIRST
 NO.   NAME      VALUE      ERROR        SIZE      DERIVATIVE
  1      A0      2278.3    2.8190     -.10390E-09    .19360E-04
  2      A1      23.469    7.0501      .16109E-09    .25818E-05
  3      A2     -503.17    22.433      .11872E-08    .25067E-05
**********
\end{verbatim}
In this output {\tt A0}, {\tt A1} and {\tt A2} are the
coefficients of the quadratic form $Y= {\tt A2}\,X^2+{\tt A1}\,X+{\tt
  A0}$, with $Y= \log L$ and $X= (M_R-80.23)$,
from which one can obtain $M_R$ in GeV,
\[
M_R-80.23={\tt 2.33E-02}\, \pm\, {\tt 3.15E-02}\, \pm\, {\tt 7.08E-03}\,,
\]
where the first is the statistical and the second 
the Monte Carlo error respectively. 

\section{Conclusions}

We have introduced the relevant formalism and described two
{\tt FORTRAN} programs to reconstruct the $W$ mass 
at LEP2 using the direct fit method introduced in ref. \cite{plb}.

\end{document}